\documentclass[aps,prd,twocolumn,floatfix,preprintnumbers,altaffilletter]{revtex4-1}
\usepackage{graphicx,url,amssymb,amsmath,longtable,rotating,color,units,wasysym,subfigure,epsfig,multirow,epstopdf,enumerate}
\usepackage{xr-hyper}
\usepackage[colorlinks,urlcolor=blue,citecolor=blue,linkcolor=blue]{hyperref}
\begin{document}

\title{Gravitational-wave memory: waveforms and phenomenology}

\author{Colm Talbot}
\author{Eric Thrane}
\author{Paul D. Lasky}
\affiliation{School of Physics and Astronomy, Monash University, Clayton, Victoria 3800, Australia}
\affiliation{OzGrav: The ARC Centre of Excellence for Gravitational-wave Discovery, Clayton, Victoria 3800, Australia}
\author{Fuhui Lin}
\affiliation{School of Physics and Astronomy, Monash University, Clayton, Victoria 3800, Australia}
\affiliation{Swarthmore College, Department of Physics and Astronomy, Swarthmore, PA 19081, USA}

\date{\today}

\begin{abstract}
The non-linear gravitational-wave memory effect is a prediction of general relativity in which test masses are permanently displaced by gravitational radiation.
We implement a method for calculating the expected memory waveform from an oscillatory gravitational-wave time series.
We use this method to explore the phenomenology of gravitational-wave memory using a numerical relativity surrogate model.
Previous methods of calculating the memory have considered only the dominant oscillatory ($\ell=2$, $m=|2|$) mode in the spherical harmonic decomposition or the post-Newtonian expansion.
We explore the contribution of higher-order modes and reveal a richer phenomenology than is apparent with $\ell=|m|=2$ modes alone.
We also consider the ``memory of the memory'' in which the memory is, itself, a source of memory, which leads to a small, $O\left(10^{-4}\right)$, correction to the memory waveform.
The method is implemented in the python package {\tt\sc GWMemory}, which is made publicly available.
\end{abstract}

\maketitle

\section{Introduction}

The non-linear (Christodoulou) gravitational-wave memory is a permanent displacement of freely-falling test masses due to the passage of gravitational waves ~\cite{Zeldovich1974,Braginsky1987,Christodoulou1991}.
This memory effect can be understood as the travelling gravitational waves themselves sourcing gravitational radiation.
Gravitational-wave memory may be detectable by advanced LIGO~\cite{aLIGO} and Virgo~\cite{AdVirgo} by considering an ensemble of detections~\cite{Lasky2016,McNeill2017}, especially if the low-frequency sensitivity can be increased~\cite{Yu2018}.
Future interferometers such as LISA, Cosmic Explorer~\cite{CosmicExplorer} and Einstein Telescope~\cite{ET} may be able to resolve the memory effect for individual binaries~\cite{Favata2009a,Yang2018}.
Gravitational-wave memory is also a target for pulsar timing arrays~\cite{vanHaasteren2010,Pshirkov2010,Seto2009,Cordes2012,Wang2015,Arzoumanian2015}.

While extracting memory from numerical relativity simulations is possible~\cite{Pollney2011}, it is time-consuming and dependent on the waveform extraction method~\cite{Taylor2013,Bishop2016,Blackman2017b}.
The minimal waveform model (MWM)~\cite{Favata2009a,Favata2009b,Favata2010} uses analytic expressions for the memory from the inspiral, using the post-Newtonian expansion, and quasi-normal mode ringdown and incorporates uncertainty in the memory sourced during merger with a ``fudge factor''.
The MWM assumes the oscillatory emission is well-described by the $\ell=|m|=2$ spin-weighted spherical harmonic modes.
For binaries with unequal masses and/or large spins this assumption is known to break down~\cite{CalderonBustillo2016,CalderonBustillo2017}.
In this work we implement a previously suggested method of calculating the memory which avoids these issues~\cite{Wiseman1991,Thorne1992,Favata2010} and explore the phenomenology of the gravitational-wave memory from binary black holes.

For the memory sourced by the $\ell=|m|=2$ modes it is possible to choose the gauge such that the memory is entirely ``$+$'' polarized and the inclination dependence is $\delta h_{+} \propto \sin^2\iota \left( 17 + \cos^2\iota \right)$.
The binary inclination, $\iota$, is the angle between the angular momentum vector of the binary and the line-of-sight between the binary and the observer.
We make this choice of gauge throughout to emphasize the deviation from the behaviour when including additional oscillatory modes.
In this paper, we demonstrate that including additional, ``higher-order'', modes in the calculation of the gravitational-wave memory leads to $O(10\%)$ corrections to the predicted strain and a richer phenomenology of gravitational-wave memory than previously believed~\footnote{The impact of higher-order oscillatory modes on the memory for non-spinning binaries is also considered in a Masters thesis by Goran Dojcinoski~\cite{PrivateCommunicationFavata}}.

Since the memory effect is sourced by gravitational radiation, the memory itself contributes to a higher-order memory effect that we call ``memory of the memory''.
We iteratively include higher-order memory terms and demonstrate that each memory order is suppressed by a factor of $\sim100$ with respect to the previous order.

Measuring gravitational-wave memory will allow new tests of general relativity and alternative theories of gravity.
For example, massive graviton theories predict a memory amplitude which is dependent on the mass of the graviton and discretely different from general relativity~\cite{Kilicarslan2018}.
Additionally, the memory effect is significantly reduced in spacetimes with more than four non-compactified dimensions~\cite{Hollands2017,Satishchandran2018,Garfinkle2018}.
Recently, it has been suggested that the inclination dependence of the memory could be used as a test of general relativity~\cite{Yang2018}.
Given that in this work we demonstrate that including higher-order oscillatory modes changes the inclination dependence, care should be taken to avoid false detection of deviations from general relativity.
Indeed, failing to consider higher-order oscillatory modes has been shown to lead to similar false detections of deviation~\cite{Pang2018}.

The remainder of the paper is structured as follows.
In the following section, we describe a method by which the gravitational-wave memory can be computed from an arbitrary spherical harmonic decomposed time-domain gravitational waveform.
We then explore the phenomenology of the gravitational-wave memory describing how the ${(\ell, m})$ content of the oscillatory waveform affects the ${(\ell, m})$ content of the memory.
After this, we consider the memory of the memory and demonstrate that the higher-order memory terms are strongly suppressed.
Finally, we present some closing thoughts.

\section{Calculating Gravitational-Wave Memory}\label{method}
The non-linear memory sourced by  gravitational waves can be expressed as an integral of the quadrupole moment of the gravitational-wave flux~\cite{Wiseman1991,Thorne1992,Favata2010}
\begin{equation}
\delta h^{TT}_{jk}(T_R, \Omega)=\frac{4G}{Rc^4}\int_{-\infty}^{T_R}dt \int_{S^{2}} d\Omega^{\prime} \frac{dE}{dt d\Omega^{\prime}} \left[ \frac{n_jn_k}{1-n^{l}N_{l}} \right]^{TT}.
\label{eq:memory}
\end{equation}
Here, $n(\Omega')$ is a unit vector, $N(\Omega)$ is the unit line-of-sight vector drawn from the observer at Earth to the source and the energy flux is
\begin{equation}
\frac{dE}{dt d\Omega}=  \frac{R^2c^3}{16\pi G}
\left|\dot{h}\left(t, \Omega\right)\right|^2,
\label{eq:flux}
\end{equation}
where $\dot{h}\equiv dh / dt$ and $h$ is the gravitational-wave strain.
We use Einstein summation convention throughout.
The angles $\Omega=(\iota, \phi)$ are the inclination and a reference phase the source (typically the phase at coalescence for compact binaries), $T_{R}$ is the retarded time, $\Omega^{\prime}$ describes a sphere centered on the source with a radius $R$, the distance between the source and the observer, and $TT$ denotes the transverse-traceless gauge.

We project onto the polarization basis by contracting with the polarization tensors, $e_{+}^{ij}$, $e_{\times}^{ij}$~\cite{Anderson2001}
\begin{equation}
\delta h = \delta h_{+} - i \delta h_{\times} = \frac{1}{2} \delta h^{TT}_{jk} (e_{+}^{jk} - i e_{\times}^{jk}).
\end{equation}

It is convenient to project the gravitational-wave strain $h(t,\Omega)$ onto a basis of spin-weighted spherical harmonics,
\begin{equation}
h(t,\Omega) = h_{+}(t,\Omega) - i h_{\times}(t,\Omega) = h_{\ell m}(t) _{-2}Y_{\ell m}(\Omega).
\label{eq:h_proj}
\end{equation}
This allows us to separate the time-dependence from the angular dependence using the same basis that is regularly used for numerical relativity waveform extraction.

Substituting Equations~\ref{eq:flux} and~\ref{eq:h_proj} into Equation~\ref{eq:memory}, we separate the time and angular integrals
\begin{equation}
\delta h(T_R, \Omega)=\frac{R}{4\pi c} H_{\ell_1 \ell_2 m_1 m_2}(T_R) \Lambda_{\ell_1 \ell_2 m_1 m_2}(\Omega),
\label{eq:separated}
\end{equation}
where we have defined
\begin{equation}
H_{\ell_1 \ell_2 m_1 m_2}(-\infty, T_R) \equiv \int^{T_R}_{-\infty} dt \dot{h}_{\ell_1 m_1}(t) \dot{\bar{h}}_{\ell_2 m_2}(t),
\end{equation}
\begin{equation}
\begin{split}
\Lambda_{\ell_1 \ell_2 m_1 m_2}(\Omega) &\equiv \frac{1}{2} (e_{+}^{jk} - i e_{\times}^{jk})  \, \times \\
\int_{S^{2}} &d\Omega' _{-2}Y_{\ell_1 m_1}(\Omega') _{-2}\bar{Y}_{\ell_2 m_2}(\Omega') \left[ \frac{n_jn_k}{1-n_{l}N_{l}} \right]^{TT} .
\label{eq:lambda}
\end{split}
\end{equation}
Overbars denote the complex conjugate.
We note that $\delta h \propto 1 / R$ as $H_{\ell_1 \ell_2 m_1 m_2} \propto 1 / R^2$.

We perform one more projection of $\Lambda_{\ell_1 \ell_2 m_1 m_2}$ onto the basis of spin-weighted spherical harmonics to facilitate combination of the oscillatory and memory waveforms,
\begin{equation}
\begin{split}
        &\Gamma_{\ell m}^{\ell_{1} \ell_{2} m_{1} m_{2}} \equiv \int_{S^{2}} d\Omega \Lambda_{\ell_{1} \ell_{2} m_{1} m_{2}} {}_{-2}\bar{Y}_{\ell m} \\
        &= 2 \pi \int_{-1}^{1} d\cos\iota \Lambda_{\ell_{1} \ell_{2} m_{1} m_{2}}(\iota, 0) {}_{-2}\bar{Y}_{\ell m_{1} - m_{2}}(\iota, 0) ,
\end{split}
\end{equation}
where we have used the fact that $\Lambda_{\ell_1 \ell_2 m_1 m_2} \propto e^{i(m_1-m_2)\phi}$ to perform the integral over $\phi$ and evaluate the $\iota$ integral at $\phi=0$.
The variable $\Gamma$ is a purely geometric factor, which we can think of as the coupling constant linking oscillatory ``input'' modes $(l_1, m_1, l_2, m_2)$ to memory ``output'' mode $(\ell, m)$.
The coefficients $\Gamma_{\ell m}^{\ell_1 \ell_2 m_1 m_2}$ are independent of the oscillatory waveform and so can be computed in advance to speed up evaluation at runtime.
It is then necessary only to compute $H_{\ell_1 \ell_2 m_1 m_2}$ and look up the relevant $\Gamma_{\ell m}^{\ell_1 \ell_2 m_1 m_2}$.

The memory accumulates over the entire lifetime of the binary, however, we are only interested here in the memory sourced from the final moments of the inspiral, merger and ringdown.
Thus, we define the lower-limit of the time integral $T_{0}$ to be the time at which the binary enters the sensitive band of our detector, usually taken to be $\unit{20}{\text{Hz}}$ for current detectors. 
Finally, we obtain
\begin{equation}
\delta h_{\ell m} = \frac{R}{4\pi c} \Gamma_{\ell m}^{\ell_1 \ell_2 m_1 m_2}(\Omega) H_{\ell_1 \ell_2 m_1 m_2}(T_{0}, T_{R}).
\label{eq:output_modes}
\end{equation}

\section{Memory Phenomenology}\label{phenomenology}

\subsection{Importance of Higher-Order Modes}\label{mode_comparison}

\begin{figure*}
\includegraphics[width=\linewidth]{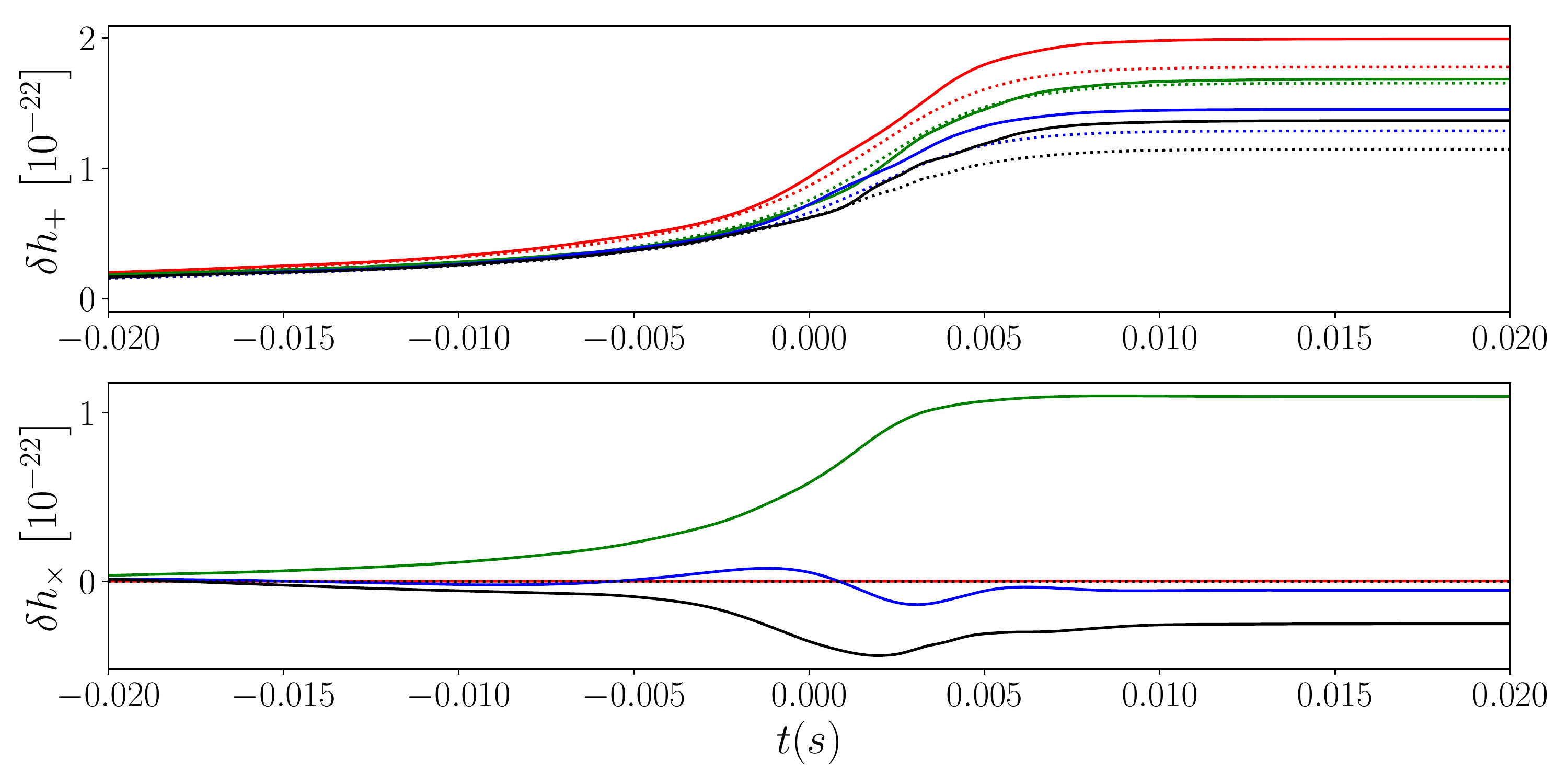}
\caption{
Including higher-order oscillatory modes significantly affects the predicted memory.
Comparison of the $+$ (top panel) and $\times$ (bottom panel) polarizations of the memory time series when using only the $\ell=|m|=2$ oscillatory modes (dotted) and when using all modes with $\ell\leq4$ (solid).
The colors are for binaries as follows: red is equal-mass ($q=1$) and non-spinning ($S_1=S_2=\vec{0}$), green is equal-mass with precessing spins ($S_{||}=0$, $S_{\bot}=0.8$), blue is unequal-mass and non-spinning, black is unequal-mass ($q \equiv m_1 / m_2 = 2$) with precessing spins.
In all cases, the late-time memory is different by $O(10\%)$ compared with the $\ell=|m|=2$ only case and is larger for large mass ratios and large, precessing, spins.
For non-spinning binaries, this is due to the excitation of higher-order modes during merger and ringdown.
Ignoring the higher-order modes completely removes the predicted $\times$ polarized memory.
The systems shown are edge-on ($\iota=\pi/2$, $\phi=0$) with total mass, $M = \unit{60}{M_\odot}$, at a luminosity distance, $D_L = \unit{400}{\mathrm{Mpc}}$.
\label{fig:comparison}
}
\end{figure*}

Previous studies of the gravitational-wave memory effect from compact binary coalescences have considered only memory sourced by the dominant, $\ell=|m|=2$ mode of the oscillatory waveform.
As mentioned above, in this case the angular dependence is given by~\cite{Favata2009a}
\begin{equation}
\delta h_{+} \propto \sin^2\iota \left( 17 + \cos^2\iota \right), \, \delta h_{\times} = 0 .
\end{equation}
This relation breaks down when additional modes are included and the angular dependence of the memory will depend on the relative size of the oscillatory spherical harmonic modes.

For our study, we use a numerical-relativity surrogate model, {\tt NRSur7dq2}~\cite{Blackman2017b}.
This model approximates the strain for all spin-weighted spherical harmonic modes with $2\leq\ell\leq4$ and is valid for mass ratios $1\leq q \equiv m_1 / m_2 \leq 2$ and dimensionless spin magnitudes up to $0.8$.
For all figures we choose a binary with a total mass of $\unit{60}{M_\odot}$ at a luminosity distance of $\unit{400}{\mathrm{Mpc}}$ with binary inclination and polarization $\iota=\pi/2$, $\phi=0$, unless otherwise stated.
We begin the integration $\unit{0.08}{\mathrm{s}}$ before the merger.

The importance of including the higher-order modes in the calculation of memory is demonstrated in Fig.~\ref{fig:comparison}.
We show the expected memory signal when considering only the $\ell=|m|=2$ oscillatory modes and when using all modes with $\ell\le4$.
We consider both non-spinning binaries and binaries with significant in-plane spins.
The in-plane spins lead to precession of the orbital plane of the binary and have a larger contribution from higher-order oscillatory modes.

We can see that even in the case of an equal-mass non-spinning binary, including the higher-order modes leads to an $O(10\%)$ change in the predicted memory signal.
This is due to the excitation of higher-order modes during the merger and ringdown portions of the coalescence.
This effect is even more pronounced for precessing, unequal-mass, binaries.
We observe that all of the considered systems other than the equal mass, non-spinning binary have a non-zero $\times$ component of the memory when the higher-order modes are included whereas the $\ell=|m|=2$ memory is entirely plus polarized.

\subsection{Mode Decomposition of the Oscillatory Waveform}\label{input_modes}

\begin{figure*}
\includegraphics[width=\linewidth]{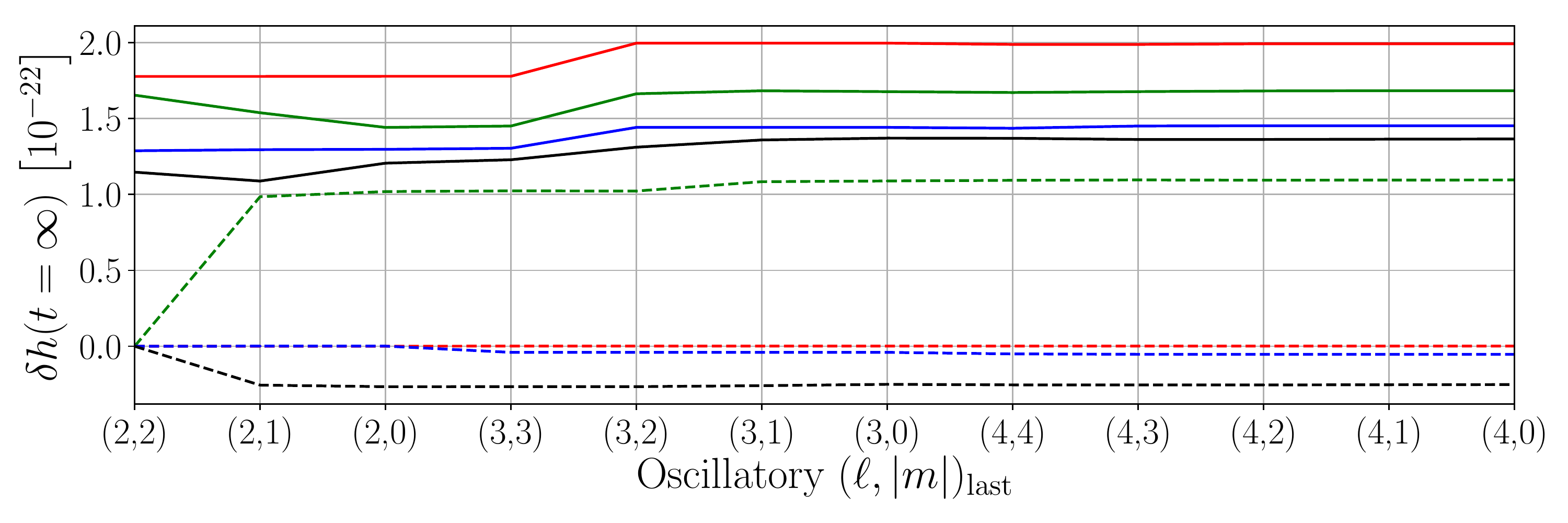}
\caption{
The late time $+$ (solid) and $\times$ polarizatons (dashed) of the memory amplitudes when including increasing numbers of modes in the oscillatory waveform (left to right).
The horizontal axis $(\ell, |m|)_\text{last}$ indicates the last two oscillatory modes included in the calculation.
The $(\ell=2,|m|=2)_\text{last}$ modes make the dominant contribution to the $+$ polarization and have no $\times$ component for all spins and mass ratios.
When the $(\ell=2$, $|m|=1)_\text{last}$ modes are added we see that there is a non-zero $\times$ polarization for spinning systems.
Including the $(\ell=3$, $|m|=2)_\text{last}$ modes has the largest effect of all the higher-order modes on the $+$ contribution to the late-time memory.
The colors are for binaries as follows: red is equal-mass ($q=1$) and non-spinning ($S_1=S_2=\vec{0}$), green is equal-mass with precessing spins ($S_{||}=0$, $S_{\bot}=0.8$), blue is unequal-mass and non-spinning, black is unequal-mass ($q \equiv m_1 / m_2 = 2$) with precessing spins.
The systems shown are edge-on ($\iota=\pi/2$, $\phi=0$) with total mass, $M = \unit{60}{M_\odot}$, at a luminosity distance, $D_L = \unit{400}{\mathrm{Mpc}}$.
We note that the memory is not necessarily maximized for edge-on systems when higher-order modes are included.
}
\label{fig:mode_build}
\end{figure*}

We now explore the effect including additional modes in the oscillatory waveform has on the final amplitude of the memory signal for the binaries in Fig.~\ref{fig:comparison}.
We consider limits on the sum in Equation~\ref{eq:separated} by progressively adding more pairs of spherical harmonic modes.
Figure~\ref{fig:mode_build} shows how the late-time non-linear memory depends on the spherical harmonic modes considered.

We see that for non-spinning binaries (red and blue curves) the most important oscillatory modes are the $\ell=2,3$, $|m|=2$.
For unequal mass binaries (blue), there is a contribution from the $\ell=|m|=3$ modes during merger, this leads to a $\times$-polarized memory component, even in the non-spinning case.
Binaries with spins in the orbital plane (green and black curves) precess, this leads to excitation of $|m|\neq2$ modes due to mode mixing~\cite{Hannam2014}.
Since there are now terms in our sum where $|m_1|\neq|m_2|$ we see a significant $\times$-polarized component in the memory.

\subsection{Mode Decomposition of the Memory Waveform}

\begin{figure*}
\includegraphics[width=\linewidth]{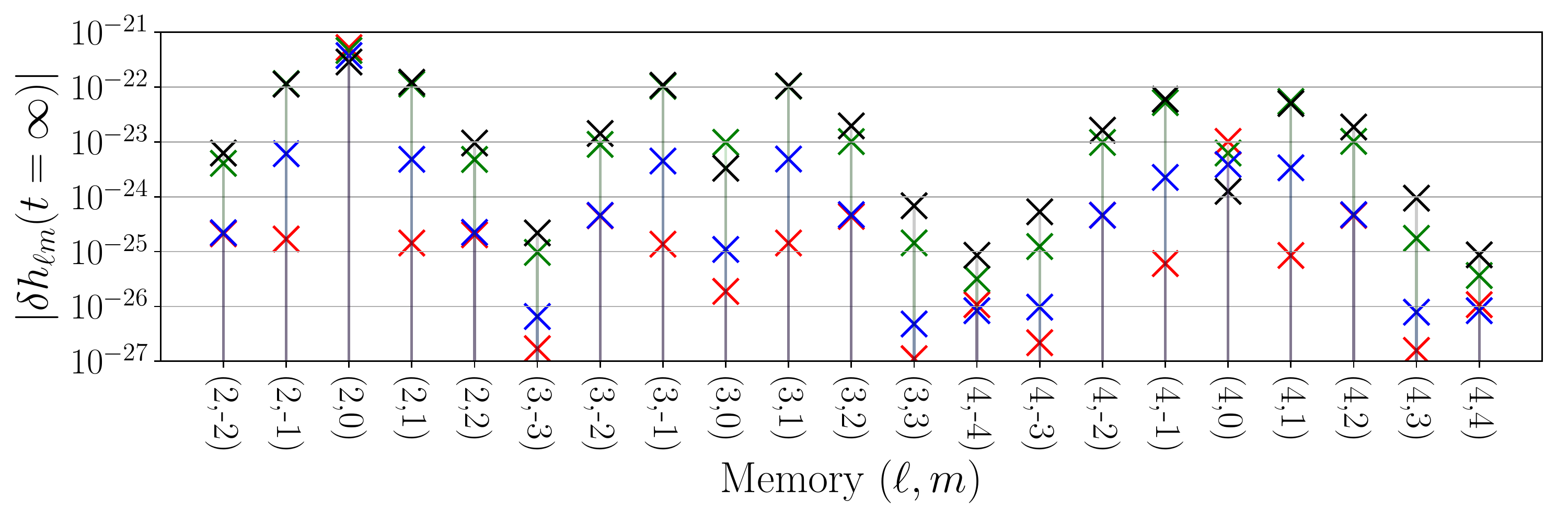}
\caption{
The spherical harmonic decomposition of the memory waveform for a range of mass ratios and spins.
The absolute value of the late-time memory is shown as a function of the $(\ell, m)$ spherical harmonic decomposition of the memory.
The dominant term is the $\ell=2$, $m=0$ mode for equal mass non-precessing binaries.
For precessing binaries, the $\times$ terms in the memory integral lead to significant azimuthal dependence of the memory.
This is seen in the $|m|=1$ modes.
The colors are for binaries as follows: red is equal-mass ($q=1$) and non-spinning ($S_1=S_2=\vec{0}$), green is equal-mass with precessing spins ($S_{||}=0$, $S_{\bot}=0.8$), blue is unequal-mass and non-spinning, black is unequal-mass ($q \equiv m_1 / m_2 = 2$) with precessing spins.
The systems shown are edge-on ($\iota=\pi/2$, $\phi=0$) with total mass, $M = \unit{60}{M_\odot}$, at a luminosity distance, $D_L = \unit{400}{\mathrm{Mpc}}$.
}
\label{fig:output_modes}
\end{figure*}

For convenience, we decompose the memory onto the basis of spin-weighted spherical harmonics.
This decomposition is given explicitly in Equation~\ref{eq:output_modes} where the $\Gamma_{\ell m}^{\ell_1 \ell_2 m_1 m_2}$ map the ``input'' oscillatory modes to the ``output'' memory modes.
Using the coefficients for $\ell = |m| = 2$ we recover the familiar $\sin^2\iota \left( 17 + \cos^2\iota \right)$ dependence.

We use the $\Gamma$ coefficients to decompose the memory onto this basis for the mass ratios and spins considered in Fig.~\ref{fig:comparison}.
The angular spectral content of these memory waveforms is shown in Fig.~\ref{fig:output_modes}.
We see that the dominant term is the $\ell=2$, $m=0$ mode in all cases.
Other modes are more important for higher mass ratios and binaries with large misaligned spins.
For precessing sources (green/black) the $|m|=1$ memory modes are nearly as large as the $m=0$ modes.
While the $m=0$ contributions to the memory decay rapidly with increasing $\ell$, the $|m|>0$ modes converge more slowly.
Therefore, it may be necessary to go consider $\ell>4$ modes to ensure waveform fidelity at the sub-percent level. 

\begin{figure*}
\includegraphics[width=\linewidth]{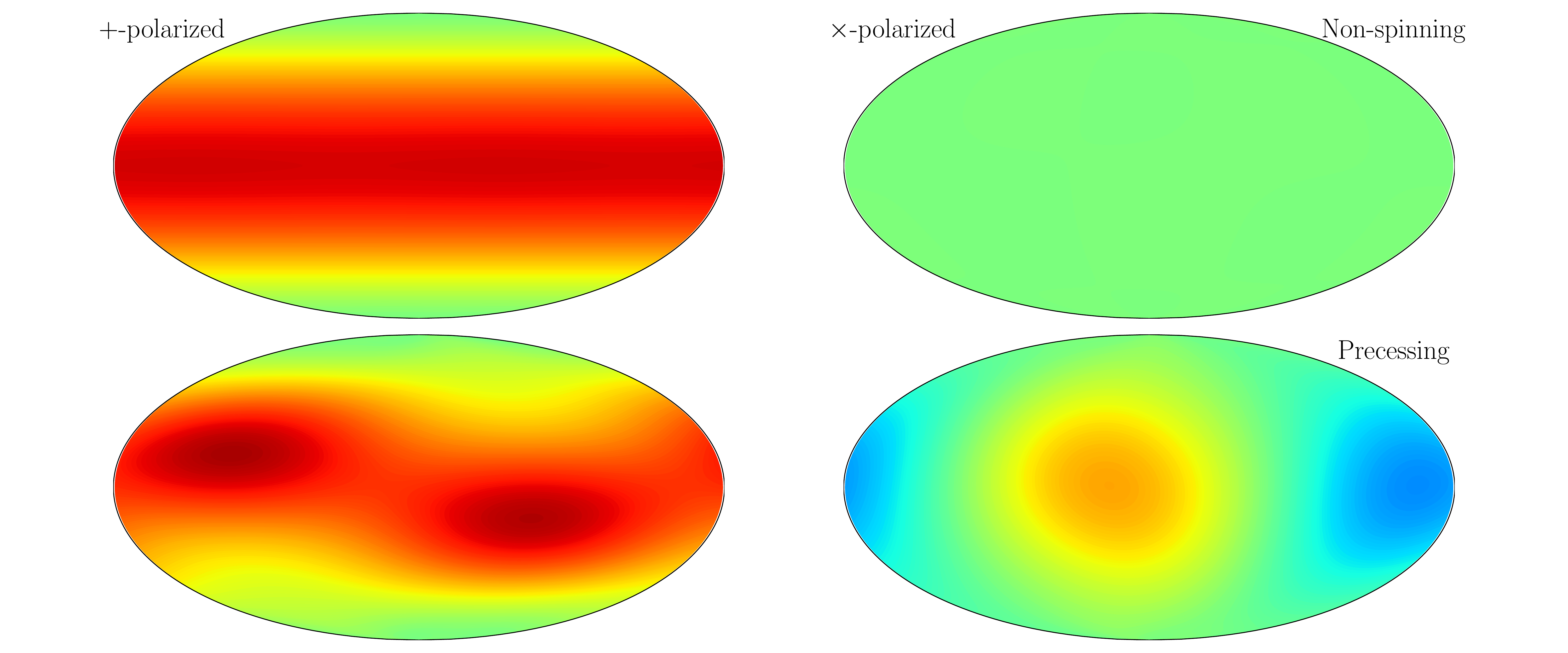}
\caption{
Angular dependence of the late-time $+$-(left) and $\times$-(right) polarized memory strain as a function of orientation angles $\iota$ (polar) and $\phi$ (azimuth).
The top panels show the late-time memory for a non-spinning equal-mass binary and the bottom panels the late-time memory for a precessing equal-mass binary.
The top panel follows the analytic expression given an oscillatory waveform containing only the $\ell=|m|=2$ mode, $\delta h_{+} \propto \sin^2\iota(17 + \cos^2\iota)$, $\delta h_{\times}=0$.
The bottom panel demonstrates how precessing systems give rise to a more complex memory structure.
}
\label{fig:maps}
\end{figure*}

Figure~\ref{fig:maps} shows the angular dependence of the late-time memory as a function of binary inclination (polar) and polarization (azimuth) for an equal mass binary.
We consider two cases: non-spinning (top panels) and precessing (bottom).
The $|m|=1$ of the memory can be seen in the precessing case.
We also draw the reader's attention to the non-vanishing $\times$ polarized memory for the precessing binary, in contrast to the non-spinning case.
We note that the orientation dependence is a function of time as different memory modes grow at different rates, which is the cause of the structure in the memory time-series in Fig.~\ref{fig:comparison} for precessing systems.

\begin{figure*}
\includegraphics[width=\linewidth]{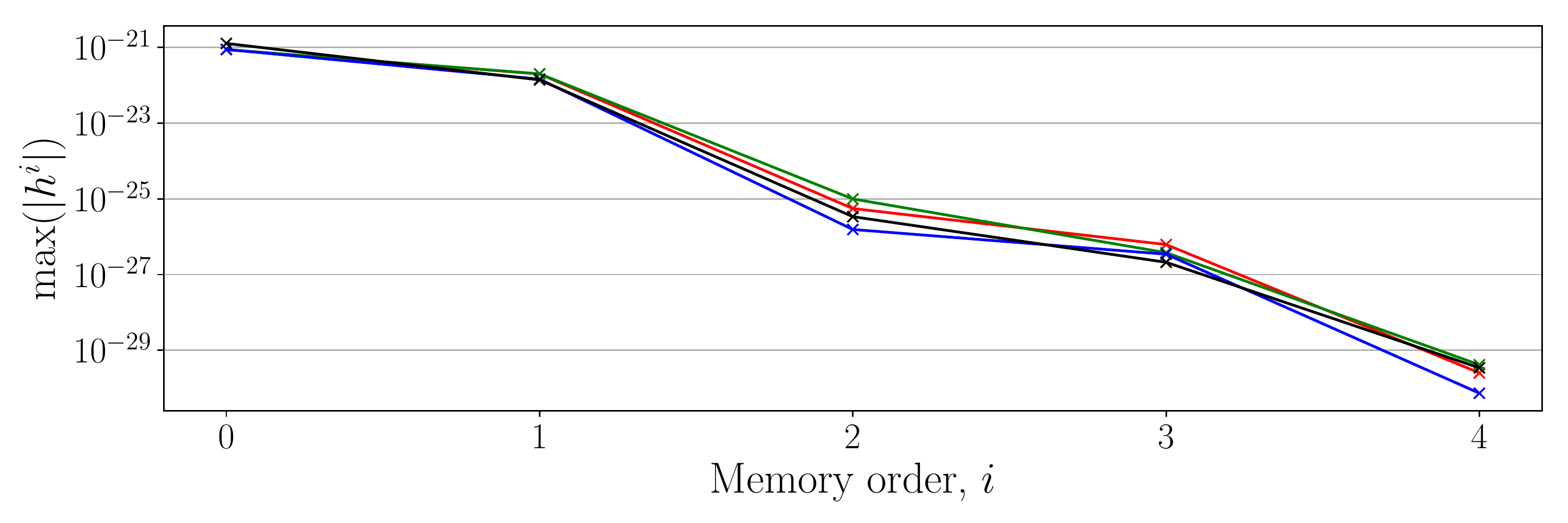}
\caption{
The maximum of the absolute value of the contribution to the memory entering at the $i$th iterative order for a range of mass ratios and spins.
The peak of the oscillatory waveform corresponds to $i=0$.
The first-order memory is $i=1$, we note that this is of the same order as the peak oscillatory strain.
Each successive order of the memory is then on average two orders of magnitude smaller than the previous.
The colors are for binaries as follows: red is equal-mass ($q=1$) and non-spinning ($S_1=S_2=\vec{0}$), green is equal-mass with precessing spins ($S_{||}=0$, $S_{\bot}=0.8$), blue is unequal-mass and non-spinning, black is unequal-mass ($q \equiv m_1 / m_2 = 2$) with precessing spins.
The systems shown are edge-on ($\iota=\pi/2$, $\phi=0$) with total mass, $M = \unit{60}{M_\odot}$, at a luminosity distance, $D_L = \unit{400}{\mathrm{Mpc}}$.
}
\label{fig:memory_of_memory}
\end{figure*}

\section{Memory of the Memory}\label{mom}

Since the memory is sourced by gravitational radiation, the memory itself imparts a second-order ``memory of the memory''.
To calculate this we replace $h_{\ell m}$ with $h^{\text{osc}}_{\ell m} + \delta h^{1}_{\ell m}$ in Equation~\ref{eq:h_proj}, where $\delta h^{1}_{\ell m}$ is the first-order memory.
We apply this procedure iteratively to calculate the total strain
\begin{equation}
h_{\ell m} = h^{\text{osc}}_{\ell m} + \sum_{i=1}^\infty \delta h^{i}_{\ell m},
\end{equation}
where $\delta h^{i}$ is the contribution to the memory entering at the $i$th order.

Figure~\ref{fig:memory_of_memory} shows the relative contribution of the different order memories for the systems considered previously.
Each successive order is suppressed by $\sim$ two orders of magnitude with respect to the previous order.
We do not expect these contributions to be significant for current detectors.
However, the sensitivity of future detectors may be sufficient to measure the memory of the memory.

\section{Memory Calculation Code}\label{code}

\begin{figure*}
\includegraphics[width=\linewidth]{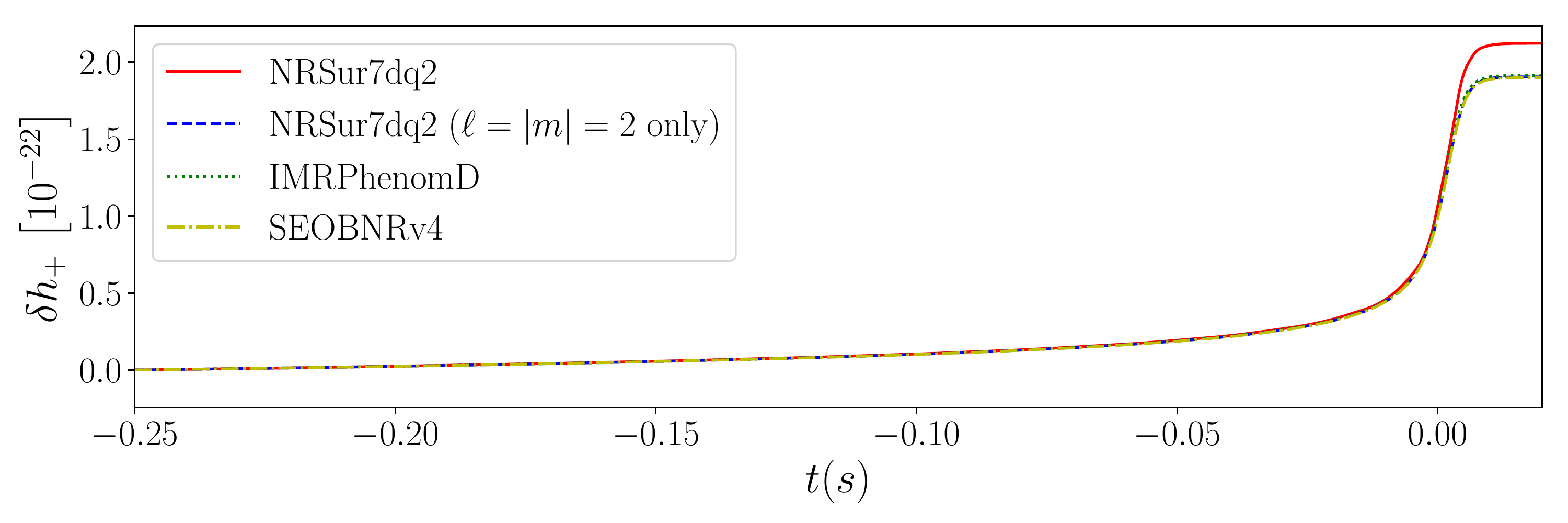}
\caption{
The plus component of the predicted memory using waveforms generated using different models.
We compare the numerical relativity surrogate (NRSur7dq2) used in the rest of the paper, an effective one-body model (SEOBNRv4) and a phenomenological model (IMRPhenomD).
The predicted memory agrees for all model when only considering the $\ell=|m|=2$ oscillatory modes.
As demonstrated above, including the higher-order oscillatory modes in the surrogate changes the predicted memory.
The system shown is edge-on ($\iota=\pi/2$, $\phi=0$), non-spinning, with total mass, $M = \unit{60}{M_\odot}$, equal mass, and at a luminosity distance, $D_L = \unit{400}{\mathrm{Mpc}}$.
}
\label{fig:waveform_comparison}
\end{figure*}

We release the Python package {\tt\sc GWMemory}~\footnote{\href{https://github.com/ColmTalbot/gwmemory}{https://github.com/ColmTalbot/gwmemory}} used in this work.
The code enables calculation of the memory from arbitrary spherical harmonic decomposed gravitational waveforms along with functionality for creating waveforms using a range of commonly-used waveform families including numerical relativity surrogates, e.g., {\tt NRSur7dq2}~\cite{Blackman2017b}, waveforms implemented in {\tt LALSuite}~\footnote{\href{https://git.ligo.org/lscsoft/lalsuite}{https://git.ligo.org/lscsoft/lalsuite}}, and numerical relativity waveforms.
Additionally, we include an implementation of the MWM~\footnote{We note that the minimal waveform model predicts a memory $\sim 20 \%$ larger than our full calculation.
We attribute this difference to the continuing development of the effective-one-body waveforms used to calibrate the MWM.}.

We have tested our waveform calculator using an aligned-spin effective one-body waveform approximant, {\tt SEOBNRv4}~\cite{Bohe2017}, a phenomenological waveform approximant, {\tt IMRPhenomD}~\cite{Khan2016}, and a numerical relativity surrogate, {\tt NRSur7dq2}~\cite{Blackman2017b}.
We find that the predicted memory does not strongly depend on the chosen oscillatory waveform family within each waveform's domain of validity, see Figure~\ref{fig:waveform_comparison}.
The surrogate model is currently limited to mass ratios $q\leq2$.
The memory for aligned-spin binaries with mass ratio $q > 2$ can be calculated using the aligned-spin waveforms available in {\tt LALSimulation}~\footnote{While precessing waveforms are available in {\tt LALSuite} the necessary decomposition into spherical harmonic modes is non-trivial and is not yet supported.}.

%
%
%
%
%

\section{Discussion}\label{discussion}
Detection of gravitational waves from binary black hole mergers allows new tests of general relativity.
In particular, we may be able to detect the gravitational-wave memory effect with current detectors~\cite{Lasky2016}.
In order to detect gravitational-wave memory using observations of merging binary black hole systems, it will be necessary to rapidly create high-fidelity frequency-domain memory waveforms for use in Bayesian parameter estimation.

The gravitational-wave memory is generally not extracted from numerical relativity simulations and is thus not modelled by the waveform approximants tuned to these simulations.
For this reason, it is necessary to calculate the expected memory waveform from the oscillatory waveform as a post-processing step.
We create a python package {\sc GWMemory} to generate the memory waveform directly from arbitrary time-domain oscillatory waveforms.

Using this code, we provide a detailed analysis of the dependence of the observed memory waveform on the spectral content of the oscillatory signal and the binary orientation~\footnote{The code used to generate the plots in this paper, along with demonstration of additional functionality, can be found at~\url{https://github.com/ColmTalbot/gwmemory/examples/GWMemory.ipynb}.}.
We find that the phenomenology of the gravitational-wave memory is richer than previously believed when sub-dominant oscillatory modes are included in the calculation of the memory.
We additionally consider the contribution of the memory waveform to a ``memory of the memory''.
While this effect is interesting from a pedagogical perspective, we find that this effect is small in all considered cases, and can be neglected with the current generation of gravitational-wave detectors.

\section*{Acknowledgements}
We thank Mark Favata, Yuri Levin, Bob Wald, Reed Essick, Juan Calderon Bustillo and Leo Stein for helpful comments and discussion.
This work is supported through Australian Research Council (ARC) Centre of Excellence CE170100004.
ET is supported through ARC Future Fellowship FT150100281.
PDL is supported through ARC Future Fellowship FT160100112 and ARC Discovery Project DP180103155.
This paper has LIGO document ID P1800183.

\bibliography{memory.bib}

\end{document}